\documentclass[12pt]{article}

\usepackage{amsmath,epsfig,epsf,psfrag,natbib}
\usepackage{amssymb}
\usepackage{amsbsy}
\usepackage{lscape}
\usepackage{color,latexsym,amsfonts,amsthm,amscd,graphicx,multirow,tabularx}
\usepackage{mathrsfs}
\usepackage{amsfonts}
\usepackage{amsmath}
\usepackage{enumitem}
\usepackage{color}
\usepackage{url}
\usepackage{mathrsfs}  
\usepackage{subfigure}
\usepackage{graphicx}
\usepackage{epstopdf}
\usepackage{hyperref}
\usepackage{bm,bbm}
\usepackage{booktabs}
\usepackage{multirow}
\usepackage{mathrsfs}
\usepackage{float}
\usepackage{ragged2e}
\usepackage{adjustbox}
\usepackage[figuresright]{rotating}
\usepackage{appendix, natbib}
\usepackage[ruled]{algorithm2e}
\usepackage{comment}
\usepackage{mathrsfs}
\usepackage{dsfont}

\usepackage{bbm}

\def\bse{\begin{eqnarray*}}
	\def\ese{\end{eqnarray*}}
\def\be{\begin{eqnarray}}
	\def\ee{\end{eqnarray}}
\def\bsq{\begin{equation*}}
	\def\esq{\end{equation*}}
\def\bq{\begin{equation}}
	\def\eq{\end{equation}}

\def\bt{{\boldsymbol\theta}}

\def\log{{\rm log}}
\def\wh{\widehat}

\def\0{{\bf 0}}
\def\wh{\widehat}

\def\Q{{\bf Q}}

\def\bb{{\boldsymbol\beta}}

\def\bSigma{{\boldsymbol\Sigma}}

\newtheorem{Theorem}{Theorem}

\newtheorem{Remark}{Remark}
\newtheorem{Proposition}{Proposition}

\def\bt{{\boldsymbol\theta}}
\def\bbeta{{\boldsymbol\beta}}

\def\bSigma{{\boldsymbol\Sigma}}

\newcommand{\bM}{ {\bf M }}

\newcommand{\bbq}{ {\bf q }}
\newcommand{\bbp}{ {\bf p }}

\newcommand{\ba}{\begin{eqnarray}}
\newcommand{\ea}{\end{eqnarray}}
\newcommand{\bas}{\begin{eqnarray*}}
\newcommand{\eas}{\end{eqnarray*}}
\newcommand{\bit}{\begin{itemize}}
\newcommand{\eit}{\end{itemize}}
\newcommand{\ben}{\begin{enumerate}}
\newcommand{\een}{\end{enumerate}}
\makeatletter
\newcommand{\Rmnum}[1]{\expandafter\@slowromancap\romannumeral #1@}
\makeatother

\usepackage{authblk}

\begin{document}
\date{}

\title{Semiparametric Joint Inference for Sensitivity and Specificity at the Youden-Optimal Cut-Off}

\author[1]{Siyan Liu}
\author[1]{Qinglong Tian}
\author[2]{Chunlin Wang\thanks{Corresponding author. Email: wangc@xmu.edu.cn}}
\author[1]{Pengfei Li}

\affil[1]{
    Department of Statistics and Actuarial Science, University of Waterloo,
 N2L 3G1, Ontario, Canada}
 \affil[2]{
    Department of Statistics and Data Science, School of Economics and Wang Yanan Institute for Studies in Economics, Xiamen University, 361005, Xiamen, China}
\renewcommand*{\Affilfont}{\small }
\renewcommand\Authands{ and }
\date{}
\maketitle

\begin{abstract}
Sensitivity and specificity evaluated at an optimal diagnostic cut-off are fundamental measures of classification accuracy when continuous biomarkers are used for disease diagnosis. Joint inference for these quantities is challenging because their estimators are evaluated at a common, data-driven threshold estimated from both diseased and healthy samples, inducing statistical dependence. Existing approaches are largely based on parametric assumptions or fully nonparametric procedures, which may be sensitive to model misspecification or lack efficiency in moderate samples.
We propose a semiparametric framework for joint inference on sensitivity and specificity at the Youden-optimal cut-off under the density ratio model. Using maximum empirical likelihood, we derive estimators of the optimal threshold and the corresponding sensitivity and specificity, and establish their joint asymptotic normality. This leads to Wald-type and range-preserving logit-transformed confidence regions. Simulation studies show that the proposed method achieves accurate coverage with improved efficiency relative to existing parametric and nonparametric alternatives across a variety of distributional settings. An analysis of COVID-19 antibody data demonstrates the practical advantages of the proposed approach for diagnostic decision-making.\\

\end{abstract}


\section{Introduction}

In medical diagnostics, a continuous biomarker is often converted into a binary decision rule by selecting a cut-off point that classifies individuals as diseased or healthy \citep{zhou2011statistical,pepe2003statistical,nakas2024roc}.
Two quantities play a central role in evaluating the classification accuracy of such a diagnostic rule:
\emph{sensitivity}, the probability of correctly identifying a diseased subject,
and \emph{specificity}, the probability of correctly identifying a healthy subject.
Both depend critically on the chosen cut-off point and are inherently linked through
the underlying biomarker distributions of diseased and healthy populations.

Among various criteria for selecting an ``optimal'' threshold, the Youden index
remains one of the most widely adopted due to its intuitive interpretation and ease of use
\citep{youden1950index,schisterman2005optimal,yuan2021semiparametric,sun2025biomarker}.
By maximizing the sum of sensitivity and specificity, the Youden index identifies the cut-off
that achieves the highest overall correct classification rate.
Let $F_0$ and $F_1$ denote the cumulative distribution functions (CDFs) of the
healthy and diseased populations, respectively.
Following \citet{yin2014joint} and \citet{hu2023statistical},
we assume that biomarker values are stochastically larger in the diseased group
and that an individual is classified as diseased when the biomarker value exceeds a threshold $x$.
Under this classification rule, the sensitivity and specificity
at the cut-off point $x$ are defined as
\begin{equation*}
\label{ge_def.spse.J}
\eta(x) = 1 - F_1(x), \qquad 
\tau(x) = F_0(x),
\end{equation*}
respectively.
The Youden index is defined as
\begin{equation}
\label{def.J}
J = \max_x \{\eta(x) + \tau(x) - 1\}
    = \max_x \{F_0(x) - F_1(x)\},
\end{equation}
and the corresponding optimal cut-off point is
\begin{equation}
\label{optim.cut}
c = \arg\max_x \{F_0(x) - F_1(x)\}.
\end{equation}
The sensitivity and specificity evaluated at this optimal cut-off are
\begin{equation*}
\label{def.spse.J}
\eta^+ = \eta(c) = 1 - F_1(c), 
\qquad 
\tau^+ = \tau(c) = F_0(c).
\end{equation*}

Although the Youden index provides a convenient scalar summary of diagnostic accuracy,
clinical decisions depend more directly on the \emph{pair}
$(\eta^+, \tau^+)$ evaluated at the selected cut-off
\citep{pepe2003statistical,nakas2024roc}.
Different biomarkers may yield the same Youden index yet exhibit substantially different trade-offs between sensitivity and specificity, underscoring the necessity of valid joint statistical inference
\citep{yin2014joint,bantis2014construction,adimari2024likelihood}.
Importantly, the optimal cut-off point $c$ is estimated from both diseased and healthy samples, so sensitivity and specificity are evaluated at a common, data-driven threshold.
Consequently, the resulting estimators of $(\eta^+, \tau^+)$ are statistically dependent, and this dependence must be properly accounted for when constructing joint confidence regions.

Existing inference procedures for $(\eta^+, \tau^+)$ largely rely on either parametric or nonparametric frameworks. Parametric methods, such as the generalized pivotal method and the Box--Cox--based delta method \citep{yin2014joint,bantis2014construction}, typically assume that the original biomarkers, or the biomarkers after a Box--Cox transformation \citep{box1964analysis}, in the healthy and diseased groups follow the same parametric distribution family. These approaches are computationally convenient and can perform well when distributional assumptions are approximately correct, but their performance may deteriorate under model misspecification. Nonparametric approaches, including smoothed bootstrap procedures \citep{yin2014joint}, avoid distributional assumptions but may yield unstable or overly conservative confidence regions when sample sizes are small or the data are highly skewed. More recently, \citet{adimari2024likelihood} proposed likelihood-type confidence regions based on empirical likelihood (EL) formulations, allowing for greater flexibility and avoiding prespecified elliptical shapes. While these methods represent an important advance, numerical studies indicate that their finite-sample performance may be unstable in certain settings, particularly when the estimated likelihood critical value becomes large or when the resulting region is poorly defined. Moreover, fully nonparametric approaches do not exploit structural similarities that often exist between healthy and diseased populations in biomedical applications, potentially resulting in inefficient inference. To date, joint inference for $(\eta^+, \tau^+)$ within a semiparametric framework that leverages such information has not been systematically developed.

To address this gap, we adopt a semiparametric modeling approach that relates the biomarker distributions in the two populations through a low-dimensional structure. In many applications, it is reasonable to assume that the diseased and healthy groups are connected through such a structural relationship rather than being completely unrelated \citep{chen2013quantile}. 
Our proposal is built on modeling the density ratio of the diseased and healthy groups, which can be constructed naturally from the logistic regression model.
Specifically, we link disease status $D$ ($D=0$ for healthy and
$D=1$ for diseased) and biomarker value $X$ through the logistic regression model:
\begin{equation}
\label{drm_logis}
P(D = 1 \mid X = x) =
\frac{\exp\{\alpha^* + \boldsymbol{\beta}^\top \mathbf{q}(x)\}}
     {1 + \exp\{\alpha^* + \boldsymbol{\beta}^\top \mathbf{q}(x)\}},
\end{equation}
where $\mathbf{q}(x)$ is a prespecified $p$-dimensional basis function, $\alpha^*$ is an
unknown scalar parameter, and $\boldsymbol{\beta}$ is a vector of unknown parameters.
Then, recall that we have $F_0(x) = P(X \le x \mid D = 0)$ and $F_1(x) = P(X \le x \mid D = 1)$. By Bayes' formula, model~\eqref{drm_logis} is equivalent to 
\begin{equation}
\label{DRM}
f_1(x) = \exp\{\alpha + \boldsymbol{\beta}^\top \mathbf{q}(x)\} \, f_0(x)
       = \exp\{\boldsymbol{\theta}^\top \mathbf{Q}(x)\} \, f_0(x),
\end{equation}
where 
$\alpha = \alpha^* - \log\{P(D = 1)/P(D = 0)\}$, 
$\boldsymbol{\theta} = (\alpha, \boldsymbol{\beta}^\top)^\top$,
$\mathbf{Q}(x) = (1, \mathbf{q}(x)^\top)^\top$,
and $f_0$ and $f_1$ denote the density
functions of $F_0$ and $F_1$, respectively. 
Model~\eqref{DRM}, known as the density ratio model (DRM; \citealp{anderson1979multivariate,qin1997goodness,qin2017biased}), quantifies the relationship between the two densities through a low-dimensional parametric component.

The DRM is semiparametric because the baseline distribution $F_0$ is left unspecified, while the density ratio is modeled parametrically. This structure~\eqref{DRM} combines interpretability with flexibility and encompasses many distributions commonly encountered in receiver operating characteristic (ROC) analysis \citep{hu2023statistical}.
For example, setting $\mathbf{q}(x) = \log(x)$ accommodates lognormal and beta families with shared shape characteristics, whereas $\mathbf{q}(x) = x$ covers the normal and exponential families with common variance or rate. The DRM has been successfully applied to ROC analysis and to inference on summary indices such as the area under the ROC curve \citep{qin2003using} and the Youden index \citep{yuan2021semiparametric}, demonstrating improved efficiency relative to fully nonparametric methods while retaining robustness to distributional misspecification. These properties make the DRM particularly appealing for joint inference on sensitivity and specificity at the Youden-optimal cut-off.

Despite these developments, joint statistical inference for $(\eta^+, \tau^+)$ under the DRM remains largely unexplored. Existing DRM-based work has primarily focused on scalar performance measures, whereas clinical interpretation and regulatory decisions often hinge on understanding the joint uncertainty of sensitivity and specificity. In this paper, we aim to develop a semiparametric framework for joint inference on $(\eta^+, \tau^+)$ under the DRM.

Our contributions are summarized as follows. First, we develop a semiparametric framework for joint inference on $(\eta^+, \tau^+)$ under the DRM and employ the maximum EL principle for estimation. We show that the optimal cut-off estimator can be obtained by solving a simple estimating equation whose solution exists within the observed data range and is unique under mild conditions, including cases where $\mathbf{q}(x)$ is a monotone scalar function of $x$.
Second, we establish the joint asymptotic distribution of the resulting estimators of $(\eta^+, \tau^+)$, thereby yielding a statistically valid Wald-type confidence region for $(\eta^+, \tau^+)$.
Third, extensive simulation studies demonstrate accurate coverage and improved efficiency across a broad range of distributional settings. An application to COVID-19 antibody data further illustrates the practical relevance of the proposed methods for diagnostic decision-making.

The remainder of this paper is organized as follows. Section~\ref{Main_results} introduces the maximum empirical likelihood estimators (MELEs) of $(\eta^+, \tau^+)$, establishes their joint asymptotic normality, and constructs confidence regions for $(\eta^+, \tau^+)$. Section~\ref{simulation} presents simulation results, and Section~\ref{real_data_analysis} contains a real-data application. Section~\ref{Discussion} concludes with a discussion of the findings. For clarity of exposition, all technical details are deferred to the Supplementary Material.


\section{Semiparametric inference procedure}
\label{Main_results}

\subsection{Problem setup and point estimation}
\label{model_set_up}

We now describe the proposed estimation procedure under the DRM. Let 
$\{X_{01}, \ldots, X_{0n_0}\} \sim F_0(x)$ and 
$\{X_{11}, \ldots, X_{1n_1}\} \sim F_1(x)$ denote two independent random samples 
from the healthy and diseased populations, respectively, with total sample size 
$n = n_0 + n_1$. Our objective is to make inference on $(\eta^+, \tau^+)$ 
based on these two samples under model \eqref{drm_logis}, or equivalently, \eqref{DRM}.

Based on the observed data, the full likelihood can be written as
\begin{equation*}
\prod_{j=1}^{n_0} f_0(X_{0j}) 
\cdot 
\prod_{j=1}^{n_1} f_1(X_{1j})
=
\prod_{i=0}^{1} \prod_{j=1}^{n_i} 
f_0(X_{ij})
\cdot
\prod_{j=1}^{n_1}
\exp\{\bt^{\top}\Q(X_{1j})\},
\end{equation*}
where the equality follows from the DRM~\eqref{DRM}. 

Following the EL principle of \cite{owen2001empirical}, 
we represent the unknown baseline distribution $F_0$ by discrete probability 
weights assigned to the pooled observations. Specifically, we set 
$p_{ij} = f_0(X_{ij})$ for $i=0,1$ and $j=1,\dots,n_i$. 
The resulting EL function is
\begin{equation*}
L_n =
\prod_{i=0}^{1} \prod_{j=1}^{n_i} 
p_{ij}
\cdot
\prod_{j=1}^{n_1}
\exp\{\bt^{\top}\Q(X_{1j})\},
\end{equation*}
where the weights $\{p_{ij}\}$ satisfy
\begin{equation}
\label{Profile_constraints}
p_{ij} \ge 0, \quad
\sum_{i=0}^{1} \sum_{j=1}^{n_i} p_{ij} = 1, \quad
\sum_{i=0}^{1} \sum_{j=1}^{n_i} 
p_{ij} \exp\{\bt^{\top}\Q(X_{ij})\} = 1.
\end{equation}
The first two constraints ensure that $F_0$ is a valid CDF, 
while the last constraint guarantees that $F_1$, defined through the DRM \eqref{DRM}, is also a valid CDF.

Let $\bbp = (p_{01}, \ldots, p_{0n_0}, p_{11}, \ldots, p_{1n_1})^\top$. 
The MELEs of $(\bt, \bbp)$, 
denoted by $(\wh\bt, \wh\bbp)$, are defined as
\[
(\wh\bt, \wh\bbp)
=
\arg\max_{\bt,\,\bbp}
L_n
\]
subject to the constraints in \eqref{Profile_constraints}. 
The numerical computation of $(\wh\bt, \wh\bbp)$ is described in Section~1 of the Supplementary Material.

Once $(\wh\bt, \wh\bbp)$ is obtained, the MELEs of $F_0$ and $F_1$ are
\[
\widehat{F}_0(x)
=
\sum_{i=0}^1 \sum_{j=1}^{n_i} \widehat{p}_{ij} \, \mathbf{1}(X_{ij} \le x),
\qquad
\widehat{F}_1(x)
=
\sum_{i=0}^1 \sum_{j=1}^{n_i}
\widehat{p}_{ij} \exp\{\widehat{\boldsymbol{\theta}}^\top Q(X_{ij})\}
\mathbf{1}(X_{ij} \le x),
\]
where $\mathbf{1}(\cdot)$ denotes the indicator function.
Note that they are estimated based on the two-sample pooled data to gain efficiency.

Recall the definition of the optimal cut-off in \eqref{optim.cut}. 
The optimal threshold $c$ satisfies
\[
f_1(c) = f_0(c),
\]
which, under the DRM~\eqref{DRM}, is equivalent to
\[
\bt^{\top} \Q(c) = 0.
\]
Replacing $\bt$ with its MELE $\wh\bt$, the estimator $\wh c$ is defined as the solution to
\begin{equation}
\label{hatc}
\wh \bt^{\top} \Q(x) 
= \wh{\alpha} + \wh{\bbeta}^\top \mathbf{q}(x) 
= 0.
\end{equation}
The corresponding plug-in MELEs of sensitivity and specificity at the optimal cut-off are
\begin{equation}
\label{spse_Estimation}
\wh{\eta}^+ = 1 - \wh F_1(\wh c),
\qquad
\wh{\tau}^+ = \wh F_0(\wh c).
\end{equation}

The following proposition establishes that $\widehat{c}$ is well defined 
and lies within the observed data range.

\begin{Proposition}
\label{prop1}
Let $X_{(1)}=\min\{X_{ij}: i=0,1,\, j=1,\ldots,n_i\}$ and 
$X_{(n)}=\max\{X_{ij}: i=0,1,\, j=1,\ldots,n_i\}$. 
Assume that $\mathbf{q}(x)$ is continuous in $x$. 
\begin{itemize}
    \item[(i)] The equation 
    $\wh{\alpha}+ \wh{\bm{\beta}}^\top \mathbf{q}(x) = 0$ 
    has at least one solution in $[X_{(1)}, X_{(n)}]$.
   
    \item[(ii)] If $\wh{\alpha} + \wh{\bm{\beta}}^\top \mathbf{q}(x)$ 
    is strictly monotone in $x$, then the equation 
    $\wh{\alpha}+ \wh{\bm{\beta}}^\top \mathbf{q}(x) = 0$  
    has a unique solution in $[X_{(1)}, X_{(n)}]$.
\end{itemize}
\end{Proposition}

Part (i) of Proposition 1 guarantees that $\wh c$ lies within the observed data range. 
In practice, $\wh c$ is computed by solving \eqref{hatc} over 
$[X_{(1)}, X_{(n)}]$ using standard one-dimensional root-finding algorithms. 
If multiple solutions arise, we select the one that maximizes 
$\wh F_0(x) - \wh F_1(x)$, in accordance with the definition of the Youden index in \eqref{def.J}. 
When $\mathbf{q}(x)$ is a monotone scalar function (e.g., $\mathbf{q}(x)=\log(x)$ or $\mathbf{q}(x)=x$) and 
$\wh{\bm{\beta}}\neq 0$, the solution of $\widehat{c}$ is unique. 
Once $\wh c$ is obtained, $(\wh\eta^+,\wh\tau^+)$ follow directly from \eqref{spse_Estimation}.

\subsection{Asymptotic  properties}
\label{joint_distribution}

 In this subsection, we establish the joint asymptotic normality of 
$(\wh{\eta}^+,\wh{\tau}^+)$. 
We begin by introducing some notation. 

Let $\rho = n_1/n$ and assume that $\rho$ is a fixed constant in $(0,1)$. 
This assumption is adopted for notational simplicity and does not affect the asymptotic results. 
We denote the true values of $\eta^+$, $\tau^+$, and the optimal threshold $c$ by 
$\eta^{+*}$, $\tau^{+*}$, and $c^*$, respectively. 
Let $\bt^*$ denote the true value of $\bt$, and define
\[
h_1(x)
=
\frac{\rho \exp\{(\bt^*)^{\top}\Q(x)\}}
{1 - \rho + \rho \exp\{(\bt^*)^{\top}\Q(x)\}}.
\]
Finally, let $\mathbf{q}'(x)$ denote the derivative of $\mathbf{q}(x)$ with respect to $x$.

The following theorem establishes the joint asymptotic normality of 
$(\wh{\eta}^+,\wh{\tau}^+)$. 
A proof is provided in Section~3 of the Supplementary Material.

\begin{Theorem}
\label{asymptotic_normality_se_sp}
Suppose that Conditions C1--C4 in the Appendix hold. 
As $n \to \infty$,
\[
\sqrt{n}
\left(
\begin{array}{c}
\wh{\eta}^+ - \eta^{+*} \\
\wh{\tau}^+ - \tau^{+*}
\end{array}
\right)
\;\longrightarrow\;
\mathrm{N}(\mathbf{0},\bSigma),
\]
in distribution, where $\mathbf{0}$ is a vector of zeros and $\bSigma = \mathbf{H}\mathbf{V}\mathbf{H}^{\top}$, with
\[
\mathbf{H}
=
\left(
\begin{array}{ccc}
\{\rho^{-1}\mathbf{B}_0 + (1-\rho)^{-1}\mathbf{B}_1\}\mathbf{B}_2^{-1} 
& 0 & -1 \\[0.1in]
(1-\rho)^{-1}(\mathbf{B}_0 - \mathbf{B}_1)\mathbf{B}_2^{-1} 
& 1 & 0
\end{array}
\right),
\]
and
\[
\mathbf{B}_0 = -\int_{-\infty}^{c^*}
h_1(x)\Q^\top(x)\, \mathrm{d}F_0(x),~~
\mathbf{B}_1 =
\frac{f_0(c^*)\Q^\top(c^*)}{\bb^{\top}\mathbf{q}'(c^*)},
~~
\mathbf{B}_2 =
\int_{-\infty}^\infty
h_1(x)\Q(x)\Q^\top(x)\, \mathrm{d}F_0(x).
\]
The explicit form of $\mathbf{V}$ is given in Lemma~S2 of the Supplementary Material.
\end{Theorem}

The proof of Theorem~\ref{asymptotic_normality_se_sp} is nontrivial because both $\wh{\eta}^+$ and $\wh{\tau}^+$ depend on $\wh{c}$, and $\wh{c}$ appears inside the indicator function. This leads to nonsmooth functionals and complicates the asymptotic analysis. Our proof combines EL theory under the DRM with empirical process techniques to address this issue; we refer to Section~3 of the Supplementary Material for details.

The asymptotic variance-covariance matrix $\bSigma$ explicitly accounts for the dependence and additional variability introduced by estimating the optimal cut-off and forms the basis for constructing joint confidence regions in the next subsection.

\subsection{Joint inference procedure}
\label{joint_inference}

In this subsection, we construct a confidence region for 
$(\tau^+,\eta^+)$. 
Based on the joint asymptotic normality established in 
Theorem~\ref{asymptotic_normality_se_sp}, 
a Wald-type confidence region can be constructed.

To estimate the variance-covariance matrix $\bSigma$, 
we replace the unknown quantities by their consistent estimators. 
Specifically, we replace 
$(\bt^*, \eta^{+*}, \tau^{+*}, F_0, F_1, c^*)$ 
with their MELEs 
$(\wh\bt, \wh\eta^{+}, \wh\tau^{+}, \wh F_0, \wh F_1, \wh c)$, 
and estimate $f_0(c^*)$ and $f_1(c^*)$ using kernel methods.

Given $\wh F_0$ and $\wh F_1$, 
the density functions $f_0$ and $f_1$ are estimated by
\[
\wh f_0(x)
=\int_{-\infty}^{\infty} K_{h_0}(x-u) \, \mathrm{d} \wh F_0(u),
\qquad
\wh f_1(x)
=\int_{-\infty}^{\infty} K_{h_1}(x-u) \, \mathrm{d} \wh F_1(u),
\]
where $K_h(x)=K(x/h)/h$ and $K(\cdot)$ is the standard normal kernel. 
The bandwidths are selected according to Silverman's rule of thumb \citep{silverman1986density}:
\[
h_0=1.06\,n^{-1/5}\min(\wh{\mathrm{IQR}}_0,\wh\sigma_0),
\qquad
h_1=1.06\,n^{-1/5}\min(\wh{\mathrm{IQR}}_1,\wh\sigma_1),
\]
where $\wh{\mathrm{IQR}}_i$ and $\wh\sigma_i^2$ denote the interquartile range 
and variance estimators based on $\wh F_i$, respectively, for $i=0,1$.

Recall that the optimal cut-off $c^*$ in \eqref{optim.cut} satisfies 
$f_0(c^*) = f_1(c^*)$. 
However, in finite samples, $\wh f_0(\wh c)$ and $\wh f_1(\wh c)$ may not coincide. 
We therefore estimate $f_0(c^*)$ and $f_1(c^*)$ by 
\[
\frac{1}{2}\left\{\wh f_0(\wh c) + \wh f_1(\wh c)\right\}.
\]

Let $\wh\bSigma$ denote the estimator of $\bSigma$ obtained by replacing 
$(\bt^*, \eta^{+*}, \tau^{+*}, F_0, F_1, c^*, f_0(c^*), f_1(c^*))$ 
with their corresponding estimators. Then it can be verified that $\wh\bSigma$ is a consistent estimator of $\bSigma$.

\begin{Theorem}
\label{consistency_of_hat_Sigma}
Under the conditions of Theorem~\ref{asymptotic_normality_se_sp}, 
$
\wh{\bSigma} \longrightarrow \bSigma
$
in probability as $n \to \infty$.
\end{Theorem}

Let 
$\boldsymbol{\mu} = (\eta^+, \tau^+)^\top$, 
$\boldsymbol{\mu}^* = (\eta^{+*}, \tau^{+*})^\top$, 
and 
$\widehat{\boldsymbol{\mu}} = (\widehat{\eta}^+, \widehat{\tau}^+)^\top$. 
Combining Theorems~\ref{asymptotic_normality_se_sp} and~\ref{consistency_of_hat_Sigma}, we obtain
\[
n(\widehat{\boldsymbol{\mu}} - \boldsymbol{\mu}^*)^\top 
\widehat{\boldsymbol{\Sigma}}^{-1}
(\widehat{\boldsymbol{\mu}} - \boldsymbol{\mu}^*)
\;\longrightarrow\; \chi^2_2,
\]
in distribution as $n \to \infty$.
Accordingly, an asymptotic $100(1-\gamma)\%$ Wald-type confidence region for $\boldsymbol{\mu}$ is
\begin{equation}
\label{EL_method}
\mathcal{C}_{\mathrm{Wald}}
=
\left\{
\boldsymbol{\mu}:\ 
n(\widehat{\boldsymbol{\mu}} - \boldsymbol{\mu})^\top 
\widehat{\boldsymbol{\Sigma}}^{-1}
(\widehat{\boldsymbol{\mu}} - \boldsymbol{\mu})
\le \chi^2_{2,\,1-\gamma}
\right\},
\end{equation}
where $\chi^2_{2,\,1-\gamma}$ denotes the $(1-\gamma)$-quantile of the chi-square distribution with two degrees of freedom.

The Wald-type confidence region \eqref{EL_method} is not range-preserving. 
Since $(\eta^+,\tau^+)\in(0,1)\times(0,1)$, 
$\mathcal{C}_{\mathrm{Wald}}$ may extend beyond the admissible parameter space, 
particularly when the true values are close to the boundary or the sample size is moderate.
To address this issue, we consider the logit transformation 
$\mathrm{logit}(x)=\log\{x/(1-x)\}$ 
and construct the confidence region based on the transformed parameters.

Let
\[
\mathbf{g}(\boldsymbol{\mu})
=
\mathrm{logit}(\boldsymbol{\mu})
=
\big(
\mathrm{logit}(\eta^+),
\mathrm{logit}(\tau^+)
\big)^\top.
\]
By the delta method,
\[
\sqrt{n}
\left\{
\mathbf{g}(\widehat{\boldsymbol{\mu}}) 
- \mathbf{g}(\boldsymbol{\mu}^*)
\right\}
\;\longrightarrow\;
\mathrm{N}\big(\mathbf{0},\boldsymbol{\Sigma}_{\mathrm{logit}}\big),
\]
in distribution as $n\to\infty$, where
\[
\boldsymbol{\Sigma}_{\mathrm{logit}}
=
\bM(\boldsymbol{\mu}^*)\,\boldsymbol{\Sigma}\,
\bM (\boldsymbol{\mu}^*)
\]
and
\[
\bM(\boldsymbol{\mu}^*)
=
\left(
\begin{array}{cc}
\{ \eta^{+*}(1-\eta^{+*}) \}^{-1} & 0 \\
0 & \{ \tau^{+*}(1-\tau^{+*}) \}^{-1}
\end{array}
\right). 
\]

Replacing $\boldsymbol{\mu}^*$ and $\boldsymbol{\Sigma}$ by their corresponding estimators $\widehat{\boldsymbol{\mu}}$ and $\wh{\bSigma}$ yields
\[
\widehat{\boldsymbol{\Sigma}}_{\mathrm{logit}}
=
\bM(\widehat{\boldsymbol{\mu}})\,
\widehat{\boldsymbol{\Sigma}}\,
\bM^\top(\widehat{\boldsymbol{\mu}}).
\]
It follows that
\[
n\left\{
\mathbf{g}(\widehat{\boldsymbol{\mu}})
- \mathbf{g}(\boldsymbol{\mu}^*)
\right\}^\top
\widehat{\boldsymbol{\Sigma}}_{\mathrm{logit}}^{-1}
\left\{
\mathbf{g}(\widehat{\boldsymbol{\mu}})
- \mathbf{g}(\boldsymbol{\mu}^*)
\right\}
\;\longrightarrow\;
\chi^2_2,
\]
in distribution as $n \to \infty$. 
Hence, the $100(1-\gamma)\%$ confidence region for $\boldsymbol{\mu}$ based on the logit transformation is
\begin{equation}
\label{EL_method_logit}
\mathcal{C}_{\mathrm{logit}}
=
\left\{
\boldsymbol{\mu}:\ 
n\left\{
\mathbf{g}(\widehat{\boldsymbol{\mu}})
- \mathbf{g}(\boldsymbol{\mu}^*)
\right\}^\top
\widehat{\boldsymbol{\Sigma}}_{\mathrm{logit}}^{-1}
\left\{
\mathbf{g}(\widehat{\boldsymbol{\mu}})
- \mathbf{g}(\boldsymbol{\mu}^*)
\right\}
\le \chi^2_{2,\,1-\gamma}
\right\}.
\end{equation}

The confidence region $\mathcal{C}_{\mathrm{logit}}$ is range-preserving by construction, 
as it is obtained through a monotone logit transformation that maps $(0,1)$ onto $\mathbb{R}$. 
Consequently, $\mathcal{C}_{\mathrm{logit}} \subset (0,1)\times(0,1)$, 
ensuring that all admissible parameter values remain within the natural parameter space. 
Our simulation studies further indicate that $\mathcal{C}_{\mathrm{logit}}$ exhibits improved coverage accuracy compared with $\mathcal{C}_{\mathrm{Wald}}$. 
Therefore, $\mathcal{C}_{\mathrm{logit}}$ is adopted in all subsequent numerical investigations.

\begin{Remark}\label{remark.gof}
The proposed estimators in \eqref{spse_Estimation} and the joint confidence region in \eqref{EL_method_logit} are derived under the DRM~\eqref{DRM}. 
In practice, it is therefore important to examine whether this semiparametric modeling strategy provides an adequate description of the data. 

For a given choice of $\bbq(x)$, one may assess the fit of \eqref{DRM} using the goodness-of-fit procedures of \citet{qin1997goodness}. 
Specifically, consider
\[
\Delta_{n0} = \sup_{x}|\widehat{F}_0(x) - \widetilde{F}_0(x)| 
\quad \text{or} \quad 
\Delta_{n1} = \sup_{x}\left| \widehat{F}_1(x) - \widetilde{F}_1(x) \right|,
\]
where
\[
\widetilde{F}_0(x) = \frac{1}{n_0}\sum_{i=1}^{n_0}\mathbbm{1}(X_{0i} \le x),
\qquad
\widetilde{F}_1(x) = \frac{1}{n_1}\sum_{i=1}^{n_1}\mathbbm{1}(X_{1i} \le x)
\]
denote the empirical distribution functions of the healthy and diseased samples, respectively. 
Bootstrap resampling was suggested to approximate the null distribution and obtain a test $p$-value.
Since $\Delta_{n0} = \rho \Delta_{n1}/(1-\rho)$, the two statistics lead to equivalent conclusions. 
In practice, it is sufficient to report one of them, for example, $\Delta_{n0}$.
\end{Remark}

\section{Simulation study}\label{simulation}
\subsection{Simulation design}

In this section, we conduct simulation studies to evaluate the finite-sample performance of the proposed point estimators in \eqref{spse_Estimation} and the confidence region in \eqref{EL_method_logit} for $(\eta^+,\tau^+)$. The proposed procedures are compared with the following five competing methods:
\begin{itemize}
\item \textbf{GPQ}: the parametric generalized pivotal method incorporating a Box--Cox transformation \citep{yin2014joint};
\item \textbf{BTATII}: a nonparametric smoothed bootstrap procedure based on the arcsine--square-root transformation, centered at the observed estimates \citep{yin2014joint};
\item \textbf{Delta BC}: the parametric delta method with a Box--Cox transformation and bootstrap-based variance estimation \citep{bantis2014construction};
\item \textbf{Emp BC}: the empirical likelihood pivot approach combined with a Box--Cox transformation \citep{adimari2024likelihood};
\item \textbf{Emp NP}: the fully nonparametric empirical likelihood method employing Gaussian kernel estimation \citep{adimari2024likelihood}.
\end{itemize}
All competing methods are primarily designed for constructing confidence regions for $(\eta^+,\tau^+)$. Each method also provides corresponding point estimators. Accordingly, our comparison assesses both point estimation accuracy and confidence region performance.

We consider three distributional settings:
\begin{enumerate}[label=(\arabic*)]
\item $f_0 \sim LN(a_0, b_0)$ and $f_1 \sim LN(a_1, b_1)$;
\item $f_0 \sim Gamma(a_0, b_0)$ and $f_1 \sim Gamma(a_1, b_1)$;
\item $f_0 \sim Beta(a_0, b_0)$ and $f_1 \sim Beta(a_1, b_1)$.
\end{enumerate}
Here, $LN(a,b)$ denotes the lognormal distribution with mean $a$ and variance $b$ on the log scale;
$Gamma(a,b)$ denotes the gamma distribution with shape parameter $a$ and rate parameter $b$ (so that the scale parameter equals $1/b$);
and $Beta(a,b)$ denotes the beta distribution with shape parameters $a$ and $b$, corresponding to $x$ and $(1-x)$, respectively.
For each distributional setting, $f_0$ is fixed, while the parameters of $f_1$ are chosen so that the resulting true Youden index, denoted by $J^*$, equals 0.3, 0.5, or 0.7, representing low, moderate, and high discriminative ability of a biomarker, respectively. The detailed parameter configurations are reported in Table~\ref{simulation_setting_CR}.
We examine four sample-size configurations,
$
(n_0, n_1) \in \{(50,50), (100,100), (50,100)$, $(100,50)\}.
$
Each scenario is replicated $1000$ times.

\begin{table}[!ht]
\centering
\caption{Parameter configurations for the three distributional settings.}
\label{simulation_setting_CR}
\setlength{\tabcolsep}{3pt} %
\renewcommand{\arraystretch}{1.05} %
\newcolumntype{C}{>{\centering\arraybackslash}X}
\begin{tabularx}{0.9\linewidth}{l CCCCCCC}
\toprule
Distribution & $J^*$ & $\eta^{+*}$ & $\tau^{+*}$ &
$a_0$ & $b_0$ &
$a_1$ & $b_1$ \\
\midrule
\multirow{3}{*}{Lognormal} 
& 0.3 & 0.65 & 0.65 & 0 & 1 & 0.77 & 1 \\
& 0.5 & 0.75 & 0.75 & 0 & 1 & 1.35 & 1 \\
& 0.7 & 0.85 & 0.85 & 0 & 1 & 2.07 & 1 \\
\midrule
\multirow{3}{*}{Gamma} 
& 0.3 & 0.697 & 0.604 & 1.5 & 1 & 2.47 & 1 \\
& 0.5 & 0.786 & 0.714 & 1.5 & 1 & 3.39 & 1 \\
& 0.7 & 0.874 & 0.826 & 1.5 & 1 & 4.81 & 1 \\
\midrule
\multirow{3}{*}{Beta} 
& 0.3 & 0.727 & 0.573 & 1.5 & 3 & 2.77 & 3 \\
& 0.5 & 0.814 & 0.686 & 1.5 & 3 & 4.25 & 3 \\
& 0.7 & 0.896 & 0.804 & 1.5 & 3 & 7.09 & 3 \\

\bottomrule
\end{tabularx}
\end{table}

For all three distributions considered, the DRM \eqref{DRM} holds with $\bbq(x)=\log(x)$, consistent with the specification suggested in the real-data analysis in Section~\ref{real_data_analysis}.
We observe that the performance patterns under unequal sample sizes, $(n_0,n_1)=(50,100)$ and $(n_0,n_1)=(100,50)$, are similar to those under equal sample sizes, $(n_0,n_1)=(50,50)$ and $(n_0,n_1)=(100,100)$. Therefore, we report only the results for equal sample sizes in the main text and present the results for unequal sample sizes in Section~4 of the Supplementary Material.

\subsection{Comparison of point estimation}

Tables~\ref{tab.eta_equal} and \ref{tab.tau_equal} report the relative bias (RB, in \%) and mean squared error (MSE, multiplied by 100) for the point estimators of $\eta^+$ and $\tau^+$, respectively; ``Our'' refers to the proposed semiparametric method.
Let $\widehat{\eta}^{+(l)}$, $l=1,\ldots,L$, denote the $l$-th simulated estimate of $\eta^+$. The RB (in percentage) and MSE are defined as
\begin{equation*}
\mathrm{RB}(\widehat{\eta}^+)
=
\frac{
\frac{1}{L}\sum_{l=1}^{L}\widehat{\eta}^{+(l)} - \eta^{+*}
}{
\eta^{+*}
}
\times 100,
\qquad
\mathrm{MSE}(\widehat{\eta}^+)
=
\frac{1}{L}
\sum_{l=1}^{L}
\left(
\widehat{\eta}^{+(l)} - \eta^{+*}
\right)^2.
\end{equation*}
The RB and MSE for the other estimators of $\eta^+$, as well as for all estimators of $\tau^+$, are computed analogously.

\begin{table}[!htt]
\centering
\small
\setlength{\tabcolsep}{0pt}
\caption{RB (\%) and MSE ($\times 100$) for the six estimators of $\eta^+$.
\label{tab.eta_equal}}
\begin{tabular*}{\textwidth}
{@{\extracolsep{\fill}} lcc@{\hspace{6pt}} cc cc cc cc cc cc @{}}
\toprule
& & & \multicolumn{2}{c}{\textbf{Our}} & \multicolumn{2}{c}{\textbf{GPQ}} & \multicolumn{2}{c}{\textbf{BTATII}} & \multicolumn{2}{c}{\textbf{Delta BC}} & \multicolumn{2}{c}{\textbf{Emp BC}} & \multicolumn{2}{c}{\textbf{Emp NP}} \\
\cmidrule(lr){4-5} \cmidrule(lr){6-7} \cmidrule(lr){8-9} \cmidrule(lr){10-11} \cmidrule(lr){12-13} \cmidrule(lr){14-15}
\text{Distribution}& {$(n_0,n_1)$}  & {$J^{*}$} & RB & MSE & RB & MSE & RB & MSE & RB & MSE & RB & MSE & RB & MSE \\
\midrule

Lognormal  & (50, 50) & 0.3  & 0.26 & 0.23 & 0.49 & 0.59 & -16.68 & 2.64 & 0.36 & 0.63 & 0.22 & 0.84 & -13.29 & 1.30 \\
       & & 0.5             & 0.15 & 0.20 & 0.13 & 0.27 & -11.44 & 1.29 & 0.43 & 0.28 & 0.34 & 0.31 & -12.20 & 1.18 \\
       & & 0.7           & 0.19 & 0.13 & -0.17& 0.14 & -9.48  & 1.07 & 0.27 & 0.13 & 0.09 & 0.17 & -11.68 & 1.22 \\
        & (100, 100) & 0.3  & 0.28 & 0.11 & 0.64 & 0.31 & -12.23 & 1.45 & 0.64 & 0.33 & 0.57 & 0.38 & -10.35 & 0.81 \\
        & & 0.5             & 0.33 & 0.09 & 0.35 & 0.14 & -8.81  & 0.75 & 0.51 & 0.14 & 0.25 & 0.16 & -9.70  & 0.74 \\
        & & 0.7             & 0.31 & 0.06 & 0.07 & 0.07 & -7.10  & 0.58 & 0.29 & 0.07 & 0.16 & 0.09 & -9.94  & 0.85 \\
\cmidrule(lr){1-15}

Gamma   & (50, 50) & 0.3   & 0.31 & 0.19 & 1.04 & 0.50 & -9.25  & 1.57 & 1.20 & 0.53 & 1.98 & 0.61 & -6.74  & 1.40 \\
        && 0.5          & 0.45 & 0.14 & 1.01 & 0.22 & -6.48  & 0.76 & 1.24 & 0.23 & 1.61 & 0.25 & -4.26  & 0.61 \\
        && 0.7            & 0.42 & 0.10 & 0.71 & 0.10 & -4.39  & 0.42 & 1.03 & 0.10 & 1.22 & 0.11 & -2.81  & 0.27 \\
        & (100, 100)& 0.3 & 0.17 & 0.10 & 0.82 & 0.28 & -7.44  & 1.06 & 0.89 & 0.29 & 1.22 & 0.32 & -4.86  & 0.98 \\
        && 0.5            & 0.15 & 0.08 & 0.82 & 0.13 & -4.84  & 0.46 & 0.93 & 0.13 & 1.03 & 0.14 & -2.81  & 0.38 \\
        && 0.7            & 0.17 & 0.05 & 0.62 & 0.06 & -2.65  & 0.21 & 0.77 & 0.06 & 0.80 & 0.06 & -1.81  & 0.17 \\
\cmidrule(lr){1-15}        
Beta    & (50, 50)& 0.3   & 0.00 & 0.17 & 0.33 & 0.44 & -3.58  & 1.31 & 0.50 & 0.46 & 1.34 & 0.39 & -3.45  & 1.50 \\
        && 0.5            & -0.09& 0.11 & 2.13 & 0.21 & -2.73  & 0.48 & 2.32 & 0.22 & 1.90 & 0.16 & -1.64  & 0.53 \\
        && 0.7            & -0.04& 0.07 & 3.01 & 0.13 & -2.31  & 0.22 & 3.20 & 0.13 & 1.65 & 0.07 & -1.10  & 0.19 \\
        & (100, 100)& 0.3 & 0.14 & 0.08 & 1.01 & 0.21 & -1.73  & 0.73 & 1.12 & 0.22 & 1.51 & 0.17 & -1.96  & 0.98 \\
        && 0.5            & 0.09 & 0.06 & 2.69 & 0.13 & -0.93  & 0.24 & 2.79 & 0.14 & 2.03 & 0.09 & -0.31  & 0.31 \\
        && 0.7            & 0.08 & 0.03 & 3.29 & 0.11 & -0.71  & 0.10 & 3.38 & 0.12 & 1.53 & 0.04 & 0.18   & 0.12 \\
\bottomrule
\end{tabular*}
\end{table}

\begin{table}[!htt]
\centering
\small
\setlength{\tabcolsep}{0pt}
\caption{RB (\%) and MSE ($\times 100$) for the six estimators of $\tau^+$.\label{tab.tau_equal}}
\begin{tabular*}{\textwidth}{@{\extracolsep{\fill}} lcc@{\hspace{6pt}} cc cc cc cc cc cc @{}}
\toprule
& & & \multicolumn{2}{c}{\textbf{Our}} & \multicolumn{2}{c}{\textbf{GPQ}} & \multicolumn{2}{c}{\textbf{BTATII}} & \multicolumn{2}{c}{\textbf{Delta BC}} & \multicolumn{2}{c}{\textbf{Emp BC}} & \multicolumn{2}{c}{\textbf{Emp NP}} \\
\cmidrule(lr){4-5} \cmidrule(lr){6-7} \cmidrule(lr){8-9} \cmidrule(lr){10-11} \cmidrule(lr){12-13} \cmidrule(lr){14-15}
\text{Distribution}  & {$(n_0,n_1)$} & {$J^{*}$} & RB & MSE & RB & MSE & RB & MSE & RB & MSE & RB & MSE & RB & MSE \\
\midrule

Lognormal & (50, 50)& 0.3    & 0.12 & 0.22 & 0.74 & 0.61 & 14.88 & 1.62 & 0.88 & 0.63 & 1.28 & 0.83 & 14.59 & 1.60 \\
        && 0.5             & 0.43 & 0.18 & 0.28 & 0.29 & 7.72  & 0.66 & 0.58 & 0.30 & 0.95 & 0.33 & 10.70 & 0.96 \\
        && 0.7             & 0.38 & 0.13 & -0.09& 0.15 & 3.80  & 0.26 & 0.34 & 0.15 & 0.74 & 0.16 & 5.56  & 0.35 \\
        & (100, 100)& 0.3  & 0.17 & 0.12 & 0.32 & 0.32 & 11.50 & 1.03 & 0.29 & 0.34 & 0.42 & 0.39 & 11.86 & 1.03 \\
        && 0.5             & 0.30 & 0.09 & 0.21 & 0.15 & 7.11  & 0.50 & 0.36 & 0.15 & 0.70 & 0.17 & 9.35  & 0.69 \\
        && 0.7             & 0.20 & 0.06 & 0.01 & 0.07 & 4.32  & 0.24 & 0.23 & 0.07 & 0.46 & 0.08 & 6.08  & 0.36 \\
\cmidrule(lr){1-15}

Gamma   & (50, 50)& 0.3   & 0.70 & 0.25 & 1.32 & 0.60 & 10.44 & 1.48 & 1.08 & 0.64 & 0.61 & 0.77 & 8.04  & 1.56 \\
        && 0.5            & 0.58 & 0.23 & 0.44 & 0.29 & 4.30  & 0.57 & 0.82 & 0.30 & 0.82 & 0.33 & 3.67  & 0.67 \\
        && 0.7            & 0.62 & 0.17 & -0.14& 0.16 & 0.01  & 0.24 & 0.39 & 0.16 & 0.51 & 0.16 & 1.19  & 0.27 \\
        & (100, 100)& 0.3 & 0.29 & 0.14 & 0.58 & 0.33 & 9.04  & 1.02 & 0.48 & 0.34 & 0.55 & 0.39 & 5.93  & 1.06 \\
        && 0.5            & 0.25 & 0.13 & 0.14 & 0.16 & 3.70  & 0.36 & 0.34 & 0.16 & 0.50 & 0.18 & 2.26  & 0.41 \\
        && 0.7            & 0.24 & 0.08 & -0.13& 0.08 & 0.65  & 0.14 & 0.15 & 0.08 & 0.26 & 0.09 & 0.87  & 0.16 \\
\cmidrule(lr){1-15}        
Beta    & (50, 50)& 0.3   & 0.13 & 0.28 & 1.84 & 0.49 & 4.69  & 1.46 & 1.56 & 0.52 & 4.43 & 0.60 & 5.43  & 1.64 \\
        && 0.5            & 0.45 & 0.28 & 0.47 & 0.26 & 0.17  & 0.58 & 0.91 & 0.27 & 4.21 & 0.34 & 1.49  & 0.67 \\
        && 0.7            & 0.67 & 0.20 & 0.43 & 0.16 & -2.23 & 0.33 & 1.10 & 0.17 & 2.54 & 0.17 & 0.25  & 0.30 \\
        & (100, 100)& 0.3 & 0.23 & 0.15 & 0.90 & 0.24 & 3.22  & 0.87 & 0.77 & 0.25 & 4.48 & 0.30 & 3.58  & 1.05 \\
        && 0.5            & 0.48 & 0.14 & 0.48 & 0.13 & 0.08  & 0.34 & 0.71 & 0.14 & 4.30 & 0.21 & 0.65  & 0.43 \\
        && 0.7            & 0.37 & 0.11 & 0.72 & 0.08 & -1.35 & 0.19 & 1.06 & 0.09 & 2.63 & 0.11 & -0.29 & 0.19 \\
\bottomrule
\end{tabular*}
\end{table}

Overall, the proposed method achieves the smallest MSE in most scenarios while maintaining consistently low RB across all distributions and sample sizes. Both GPQ and Delta BC exhibit moderately larger MSEs. Their biases become more pronounced under the beta distribution, particularly at larger values of the Youden index.
BTATII tends to show substantially higher bias and variability when the sample sizes are small, although its performance improves as the sample size increases.
Among the empirical likelihood–based approaches, Emp BC produces relatively stable results but generally yields larger MSEs than the proposed method. In contrast, Emp NP displays noticeable fluctuations in both RB and MSE across settings, indicating limited numerical stability.

\subsection{Comparison of confidence regions}

We now compare the proposed joint confidence region for 
$\boldsymbol{\mu}=(\eta^+,\tau^+)^{\top}$ 
with those obtained from the five competing methods. 
The nominal confidence level is set at 95\%. 
The performance of a confidence region is assessed by the coverage probability (CP, in \%) and the average area of the confidence region (ACR), defined as
\[
CP=\frac{1}{L} \sum_{l=1}^L 
\mathbbm{1}\bigl(\boldsymbol{\mu}^* \in \mathcal{I}^{(l)}\bigr) 
\times 100,
\qquad 
ACR=\frac{1}{L} \sum_{l=1}^L C^{(l)},
\]
where $\mathcal{I}^{(l)}$ denotes the confidence region for $\boldsymbol{\mu}$ obtained from the $l$-th simulation run, and $C^{(l)}$ represents its area.

Table~\ref{tab.combined_compact} reports the empirical CPs and ACRs for all methods, where ``Our'' refers to the proposed method. 
We note that the Emp NP method may produce numerical errors in some simulation repetitions and therefore fails to construct a confidence region for $\boldsymbol{\mu}$ in those cases. 
The number of such failures (out of 1000 repetitions) in each scenario is recorded and reported in the last column, denoted by ``\#Err''.

Across all settings, the proposed method attains coverage probabilities close to the nominal 95\% level while maintaining the smallest or near-smallest ACRs. 
The coverage of the GPQ method generally decreases as the Youden index increases, with the decline being particularly pronounced under the beta distribution.  BTATII tends to exhibit the opposite pattern in several settings, with coverage improving as the Youden index increases, although this trend is not strictly monotone across all distributions and sample sizes. 
Delta BC demonstrates intermediate performance, with coverage probabilities typically lying between those of the proposed method and the GPQ approach. However, its coverage deteriorates substantially under the beta distribution and, to a lesser extent, under the gamma distribution when the Youden index is large. This suggests that Delta BC may perform poorly when the Box--Cox transformation model is misspecified, especially at high discriminative levels. Among the two EL methods, Emp BC achieves reasonably stable nominal coverage (approximately 90\%--96\%) but produces substantially larger confidence regions than the proposed method. In contrast, Emp NP exhibits frequent computational failures. The number of failures (reported in the last column of Table~\ref{tab.combined_compact}) reaches as high as 442 out of 1000 simulations in the lognormal setting with a Youden index of 0.7.

\begin{table}[!htt]
\centering
\small
\setlength{\tabcolsep}{1.2pt} 
\caption{CP (\%) and ACR ($\times 100$) for the six 95\% confidence regions of $(\eta^+,\tau^+)$.
\label{tab.combined_compact}}
\begin{tabular*}{\textwidth}
{@{\extracolsep{\fill}} l c c@{\hspace{6pt}} cc cc cc cc cc cc c @{}}
\toprule
& & & \multicolumn{2}{c}{\textbf{Our}} & \multicolumn{2}{c}{\textbf{GPQ}} & \multicolumn{2}{c}{\textbf{BTATII}} & \multicolumn{2}{c}{\textbf{Delta BC}} & \multicolumn{2}{c}{\textbf{Emp BC}} & \multicolumn{3}{c}{\textbf{Emp NP}} \\
\cmidrule(lr){4-5} \cmidrule(lr){6-7} \cmidrule(lr){8-9} \cmidrule(lr){10-11} \cmidrule(lr){12-13} \cmidrule(lr){14-16}
\text{Distribution}& {$(n_0,n_1)$} & {$J^*$}  & CP & ACR & CP & ACR & CP & ACR & CP & ACR & CP & ACR & CP & ACR & \#Err \\
\midrule

Lognormal & (50, 50) & 0.3   & 95.2 & 3.94 & 89.9 & 9.99 & 54.9 & 6.71 & 91.7 & 9.52 & 91.0 & 15.9 & 71.3 & 20.4 & 23 \\
          && 0.5           & 95.7 & 3.31 & 87.6 & 4.60 & 69.1 & 7.33 & 91.2 & 4.98 & 93.1 & 6.85 & 63.5 & 12.1 & 77 \\
          && 0.7           & 95.4 & 2.39 & 82.7 & 1.94 & 77.8 & 6.67 & 91.7 & 2.49 & 95.8 & 4.16 & 73.3 & 7.21 & 442 \\
          & (100, 100) & 0.3 & 94.3 & 1.99 & 91.5 & 4.85 & 53.2 & 3.81 & 93.6 & 4.90 & 93.2 & 8.07 & 71.0 & 14.8 & 1 \\
          && 0.5           & 96.2 & 1.65 & 88.5 & 2.22 & 61.4 & 4.12 & 92.6 & 2.60 & 94.1 & 3.75 & 62.2 & 8.17 & 1 \\
          && 0.7           & 96.3 & 1.16 & 81.0 & 0.90 & 79.6 & 3.95 & 93.5 & 1.28 & 95.4 & 2.40 & 67.5 & 4.62 & 38 \\
\midrule

Gamma     & (50, 50) & 0.3  & 96.0 & 3.93 & 89.1 & 9.40 & 59.3 & 7.12 & 92.2 & 8.86 & 91.1 & 14.1 & 91.4 & 29.1 & 2 \\
          && 0.5          & 95.6 & 3.22 & 86.6 & 4.30 & 82.1 & 7.63 & 91.8 & 4.54 & 93.0 & 6.37 & 92.6 & 14.9 & 4 \\
          && 0.7         & 95.5 & 2.22 & 81.8 & 1.78 & 96.1 & 7.17 & 90.8 & 2.14 & 93.5 & 3.70 & 94.0 & 7.05 & 52 \\
          & (100, 100)& 0.3& 93.8 & 1.99 & 90.2 & 4.61 & 56.7 & 3.90 & 92.3 & 4.54 & 93.6 & 7.22 & 91.5 & 22.2 & 0 \\
          && 0.5          & 94.2 & 1.62 & 86.1 & 2.11 & 80.1 & 4.19 & 91.8 & 2.39 & 93.1 & 3.38 & 93.1 & 10.2 & 0 \\
          && 0.7          & 94.8 & 1.10 & 80.6 & 0.85 & 96.9 & 4.04 & 89.9 & 1.11 & 93.0 & 1.93 & 95.0 & 4.22 & 0 \\
\midrule
Beta      & (50, 50)& 0.3   & 95.3 & 3.93 & 92.8 & 9.16 & 61.7 & 6.97 & 94.0 & 8.26 & 96.1 & 10.9 & 95.5 & 31.7 & 16 \\
          && 0.5         & 95.4 & 3.24 & 83.3 & 3.75 & 87.7 & 7.48 & 90.2 & 4.00 & 95.0 & 5.67 & 95.8 & 15.1 & 33 \\
          && 0.7          & 95.6 & 2.23 & 58.1 & 1.23 & 98.5 & 6.89 & 73.2 & 1.56 & 95.1 & 3.55 & 96.1 & 6.37 & 160 \\
          & (100, 100) & 0.3 & 95.5 & 1.99 & 92.3 & 4.31 & 59.4 & 3.83 & 93.6 & 4.02 & 96.5 & 5.17 & 94.4 & 23.5 & 0 \\
          && 0.5           & 95.2 & 1.62 & 79.7 & 1.74 & 88.6 & 4.13 & 87.3 & 1.96 & 94.8 & 2.88 & 93.9 & 10.1 & 0 \\
          && 0.7          & 95.9 & 1.09 & 34.6 & 0.56 & 97.6 & 3.96 & 59.9 & 0.74 & 94.2 & 1.77 & 95.2 & 3.89 & 6 \\
\bottomrule
\end{tabular*}
\begin{flushleft}
\footnotesize
\#Err: number of times the Emp NP method fails to construct a confidence region (out of 1000 repetitions).
\end{flushleft}
\end{table}

\section{Real data analysis}\label{real_data_analysis}
We illustrate the proposed method and compare it with five competing approaches using the COVID-19 antibody dataset analyzed by \citet{Bantis_2024_Statistical}. The data consist of receptor-binding domain (RBD)–specific IgG antibody titers measured in healthcare workers following the second dose of the Pfizer–BioNTech BNT162b2 mRNA vaccine. Antibody titers (denoted by $x$), expressed in arbitrary units per milliliter (AU/mL), are available for 100 healthcare workers, including 42 infection-positive and 58 infection-negative individuals, representing the immune response approximately two weeks after the second vaccination dose. The dataset is available in the Supporting Information of \citet{Bantis_2024_Statistical}.

\begin{figure}[!ht]
\centering\includegraphics[width=0.95\textwidth]{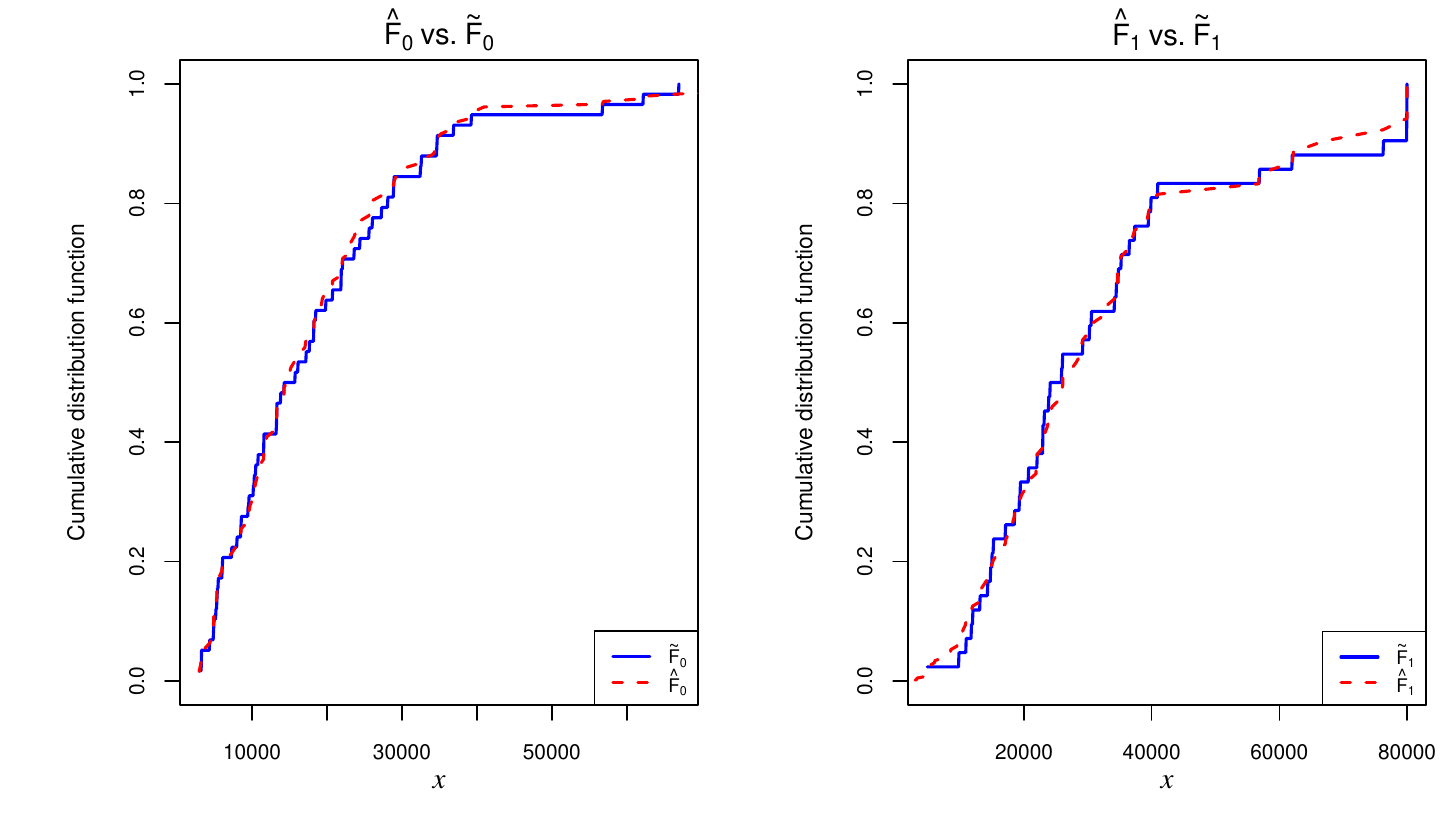}
\vspace{-0.4in}
      \caption{Comparison of the MELEs $(\widehat F_0,\widehat F_1)$ with the corresponding empirical distribution functions $(\widetilde F_0,\widetilde F_1)$ for the COVID-19 antibody data.}
      \label{Fig-real_data_empirical_dist_compare}
      \end{figure}

Without having to specify a fully parametric distribution, the only component that needs to be specified in our approach is the basis function $\bbq(x)$ in the DRM~\eqref{DRM}.
As no specific functional form for $\bbq(x)$ was prescribed a priori, we considered a range of candidate specifications, including $x$, $\log(x)$, $x^2$, $\{\log(x)\}^2$, and their combinations. Model selection criteria based on the AIC and BIC, as summarized in Table~\ref{tab:model_selection}, consistently favored the specification $\bbq(x)=\log(x)$.
To further assess model adequacy, we applied the goodness-of-fit procedure described in Remark~\ref{remark.gof} using 1000 bootstrap samples. The resulting $p$-value of 0.76 provides no evidence against the DRM under $\bbq(x)=\log(x)$.
Figure~\ref{Fig-real_data_empirical_dist_compare} compares the MELEs $(\widehat F_0,\widehat F_1)$ with the corresponding empirical distribution functions $(\widetilde F_0,\widetilde F_1)$. The close agreement between the two further supports the adequacy of the selected basis specification.

\begin{table}[!ht]
\centering
\caption{Model selection results for the COVID-19 antibody data.}
\label{tab:model_selection}
\begin{tabular}{lcc}
\toprule
$\bbq(x)$ & {AIC} & {BIC} \\
\midrule
$\log(x)$ & 907.95 & 913.16 \\
$\{\log (x)\}^2$ & 908.18 & 913.39 \\
$\log(x) + \{\log (x)\}^2$ & 909.63 & 917.45 \\
$x + \log(x)$ & 909.85 & 917.67 \\
$x + x^2 + \{\log (x)\}^2$ & 909.93 & 920.35 \\
$\log(x) + x^2$ & 909.95 & 917.77 \\
$x + \{\log (x)\}^2$ & 909.98 & 917.79 \\
$x + \log(x) + x^2$ & 910.00 & 920.42 \\
$x^2 + \{\log (x)\}^2$ & 910.15 & 917.97 \\
$\log(x) + x^2 + \{\log (x)\}^2$ & 910.44 & 920.86 \\
$x + \log(x) + \{\log (x)\}^2$ & 910.76 & 921.18 \\
$x$ & 911.41 & 916.62 \\
$x + x^2 + \log(x) + \{\log (x)\}^2$ & 911.54 & 924.57 \\
$x + x^2$ & 912.13 & 919.95 \\
$x^2$ & 914.92 & 920.13 \\
\bottomrule
\end{tabular}
\end{table}

Table~\ref{tab:se_sp_area} presents the point estimates of $(\eta^+,\tau^+)$, together with the corresponding areas ($\times 100$) of the 95\% joint confidence regions obtained from the competing methods. Among these, Emp NP yields the largest confidence region, whereas the proposed method produces the smallest area ($3.889\times 10^{-2}$). The Delta BC, Emp BC, BTATII, and GPQ procedures yield intermediate region sizes. These findings are consistent with the simulation results.

\begin{table}[!htt]
\centering
\caption{Point estimates (PEs) of $(\eta^+,\tau^+)$ and areas ($\times 100$) of the corresponding 95\% confidence regions (CRs) for the COVID-19 antibody data.}
\begin{tabular}{lccc}
\hline
Method & PE of ${\eta}^+$ & PE of ${\tau}^+$ & Area of 95\% CR\\
\hline
Our & 0.694 & 0.641 & 3.889 \\
GPQ &0.724 &0.616 & 8.427 \\
BTATII & 0.696 & 0.613 & 7.168 \\
Delta BC & 0.731 & 0.612 & 7.993 \\
Emp BC & 0.654 & 0.676  & 8.764 \\
Emp NP & 0.688 & 0.604 & 40.605 \\

\hline
\end{tabular}
\label{tab:se_sp_area}
\end{table}

\begin{figure}[!ht]
\centering\includegraphics[width=0.65\textwidth, height=0.65\textwidth]{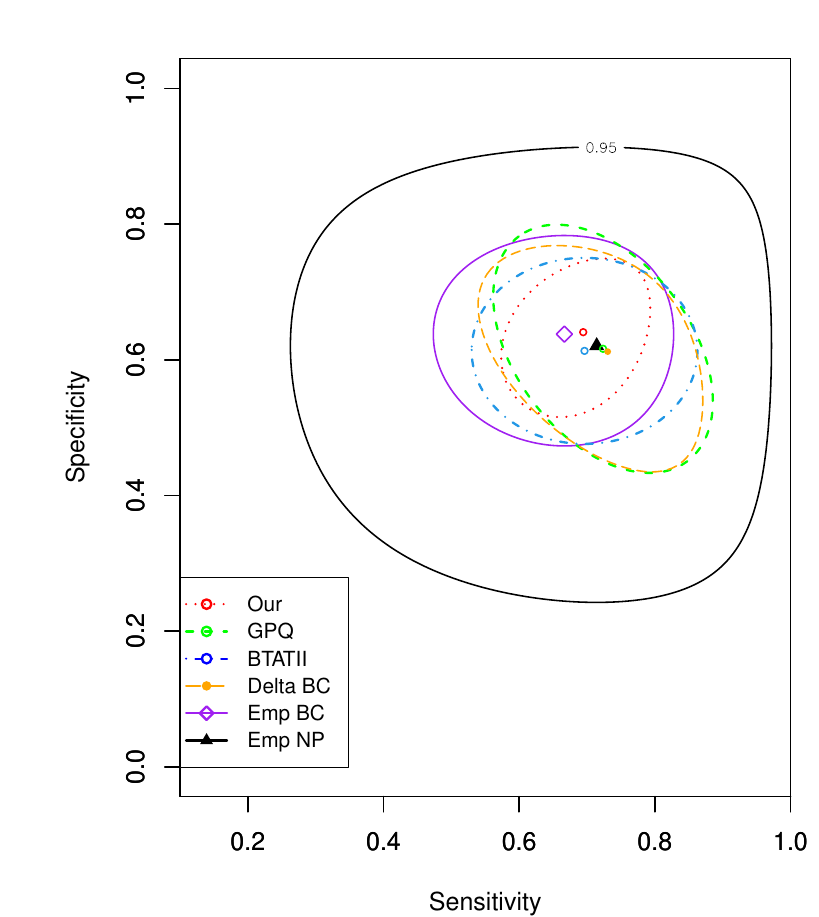}
\caption{Plots of the 95\% confidence regions for $(\eta^+,\tau^+)$ based on the COVID-19 antibody data.}
      \label{Fig-or_real_data_compare}
      \end{figure}

Figure~\ref{Fig-or_real_data_compare} displays the corresponding 95\% joint confidence regions for $(\eta^+, \tau^+)$. The region obtained from the proposed method is largely nested within those produced by the alternative approaches, indicating a higher degree of joint precision. In contrast, Emp NP generates the most diffuse region. These graphical results align with the numerical comparisons in Table~\ref{tab:se_sp_area} and further illustrate the relative efficiency of the proposed procedure.

  \section{Discussion}
  \label{Discussion} 

This paper develops semiparametric procedures for joint statistical inference on sensitivity and specificity at the Youden-optimal cut-off under the DRM. We show that the estimator of the optimal cut-off can be obtained by solving a simple estimating equation whose solution is guaranteed to lie within the observed data range and is unique when $\bbq(x)$ is a monotone scalar function. We further establish the joint asymptotic normality of the resulting estimators and construct a statistically valid confidence region for sensitivity and specificity at the Youden-optimal cut-off. Simulation studies and the analysis of the COVID-19 antibody data demonstrate the favorable finite-sample performance and practical utility of the proposed methods.

While this article focuses on the two-class setting, the proposed procedures under the DRM \eqref{DRM} can be extended to multi-class problems. In the three- and general $k$-class cases, diagnostic accuracy is characterized by ROC surfaces or manifolds and the associated vector of true class fractions (TCFs) \citep{bantis2017construction,wang2024evaluating}. Because our framework provides joint inference for a vector of accuracy measures and properly accounts for the correlation induced by data-driven cut-off estimation, the same principles can be applied to the multi-dimensional TCF parameters arising from optimal multi-threshold selection. We leave the development of these extensions to future research.

\appendix
\begin{appendices}
\section{Regularity conditions}

The asymptotic results in Theorems \ref{asymptotic_normality_se_sp} and \ref{consistency_of_hat_Sigma} are established under the following regularity conditions.

\begin{itemize}

\item[C1.] The total sample size $n = n_0 + n_1 \to \infty$, and 
$\rho = n_1 / n$ converges to a constant in $(0,1)$.

\item[C2.] 
The two distribution functions $F_0$ and $F_1$ satisfy the DRM \eqref{DRM}. In addition,
\[
\mathbf{B}_2 =
\int_{-\infty}^{\infty}
h_1(x)\mathbf{Q}(x)\mathbf{Q}^\top(x)\, \mathrm{d}F_0(x)
\]
is positive definite. Moreover, for $\boldsymbol{\theta}$ in a neighborhood of $\boldsymbol{\theta}^*$,
\[
\int_{-\infty}^{\infty} 
\exp \left\{\boldsymbol{\theta}^{\top} \mathbf{Q}(x)\right\} 
\, \mathrm{d} F_0(x) < \infty.
\]

\item[C3.] 
For any $\epsilon > 0$, 
\[
J_\epsilon = \sup_{|x - c^*| > \epsilon}
\{F_0(x) - F_1(x)\} < J^* .
\]

\item[C4.] 
For $i=0,1$, the distribution function $F_i$ is twice continuously differentiable in a neighborhood of $c^*$, with first and second derivatives denoted by $f_i$ and $f_i'$, respectively. 
Furthermore, $f_0(c^*) = f_1(c^*) > 0$ and 
$f_0'(c^*) < f_1'(c^*)$.

\end{itemize}

Condition C1 ensures that $n_0$ and $n_1$ diverge at the same rate as the total sample size. 
Condition C2 guarantees the identifiability of the DRM parameters and, together with C1, ensures the asymptotic normality of $\widehat{\boldsymbol{\theta}}$. 
Conditions C3 and C4 ensure the uniqueness and identifiability of the Youden-optimal cut-off $c^*$.

 \end{appendices}

\bibliographystyle{natbib}
\bibliography{references}

\end{document}